\documentstyle[12pt,psfig]{article} %available in the standard LaTex
\normalbaselines
\global\arraycolsep=2pt

\begin{document}
\bibliographystyle{unsrt}

%\vbox {\vspace{4mm}} %Leave space at the top on the first page.
\begin{center}
{\large \hfill OKHEP-94-01}
\\[2mm]
\end{center}
\begin{center}
{\large \bf FINITE-ELEMENT TIME EVOLUTION OPERATOR\\[2mm]
FOR THE ANHARMONIC OSCILLATOR}\footnote{Contribution to
Harmonic Oscillators II, Cocoyoc, Mexico, March 23-25, 1994}\\[7mm]
Kimball A. Milton\\
{\it Department of Physics and Astronomy\\
The University of Oklahoma, Norman OK 73019, USA}\\[5mm]
\end{center}

\vspace{2mm}
\begin{abstract}
The finite-element approach to lattice field theory is both highly
accurate (relative errors $\sim1/N^2$, where $N$ is the number of
lattice
points) and exactly unitary (in the sense that canonical commutation
relations
are exactly preserved at the lattice sites).  In this talk I
construct matrix
elements for dynamical variables and for the time evolution operator
for the
anharmonic oscillator, for which the continuum Hamiltonian is
$H=p^2/2
+\lambda q^4/4.$  Construction of such matrix elements does not
require
solving the implicit equations of motion.  Low order approximations
turn
out to be extremely accurate.  For example, the matrix element of the
time evolution operator in the harmonic oscillator ground state gives
a result for the anharmonic oscillator ground state energy accurate
to better
than 1\%, while a two-state approximation reduces the error to less
than 0.1\%.
\end{abstract}
\section{Introduction}

For over a decade now, the finite-element method has been developed
for
application to quantum systems.  (For a review of the program see
\cite{review}.)  The essence of the approach is to put the Heisenberg
equations of motion for the quantum system on a Minkowski space-time
lattice
in such a way as to preserve exactly the canonical commutation
relations
at each lattice site. Doing so  corresponds precisely to the
classical
finite-element prescription of requiring continuity at the lattice
sites
while imposing the equations of motion at the Gaussian knots, a
prescription
chosen to minimize numerical error.  We have applied this technique
to
examples in quantum mechanics and to quantum field theories in two
and four
space-time dimensions.  In particular, recent work has concentrated
on
Abelian and non-Abelian gauge theories \cite{sing,mmsb,mmsb2}.

Because it is the equations of motion that are discretized, a lattice
Lagrangian does not exist in Minkowski space.  This is because the
equations of motion are in general nonlocal, involving fields at all
previous (but not later) times.  Similarly, a lattice Hamiltonian
does
not exist, in the sense of an operator from which the equations of
motion
can be derived.

However, because the formulation is  unitary, a unitary
time-evolution
operator must exist which carries fields from one lattice time to the
next.
For linear finite elements this operator in quantum mechanics
 has been explicitly constructed
\cite{dtqm}.  Construction of this operator requires solving the
equations
of motion, which are implicit.  Therefore, it is most useful, and
perhaps
surprising, that when  matrix elements of the time
evolution operator are constructed in a harmonic oscillator basis,
they do not require the solution
of the equations of motion \cite{bss}.  Although these general
formulas were
derived some years ago, it seems they have not been exploited.
My purpose here is to study, in a simple context, the matrix elements
of the evolution operator, and see how accurately spectral
information
may be extracted.  My goal, of course, is to apply similar techniques
in gauge theories, for example, to study chiral symmetry breaking in
QCD.

\section{Review of the Finite-Element Method}

Let us consider a quantum mechanical system with one degree of
freedom
governed by the continuum Hamiltonian
\begin{equation}
H={p^2\over2}+V(q),
\label{ham}
\end{equation}
from which follow the Heisenberg equations
\begin{equation}
\dot p=-V'(q),\quad \dot q=p.
\label{eqm}
\end{equation}
These equations are to be solved subject to the initial condition
\begin{equation}
[q(0),p(0)]=i.
\end{equation}
It immediately follows from (\ref{eqm}) that the same relation holds
at
any later time
\begin{equation}
[q(t),p(t)]=i.
\end{equation}

Now suppose we introduce a time lattice by subdividing the interval
$(0,T)$ into $N$ subintervals each of length $h$.  On each
subinterval
(``finite element'') we express the dynamical variables as $r$th
degree
polynomials
\begin{equation}
p(t)=\sum_{k=0}^r a_k(t/h)^k,\quad q(t)=\sum_{k=0}^r b_k(t/h)^k,
\end{equation}
where $t$ is a local variable ranging from $0$ to $h$.  We determine
the
$2(r+1)$ operator coefficients $a_k$, $b_k$, as follows:
\begin{enumerate}
\item On the first finite element let
\begin{equation}
a_0=p_0=p(0),\quad b_0=q_0=q(0).
\end{equation}
\item Impose the equations of motion (\ref{eqm}) at $r$ points within
the
finite element, at $\alpha_i h$, $i=1,2,\dots,r$, where
$0<\alpha_1<\alpha_2<
\cdots<\alpha_r<1$.
  This then gives
\begin{equation}
p(h)\approx p_1=\sum_{k=0}^r a_k,\quad q(k)\approx q_1=\sum_{k=0}^r
b_k.
\end{equation}
\item Proceed to the next finite element by requiring continuity (but
not
continuity of derivatives) at the lattice sites, that is, on the
second
finite element, set
\begin{equation}
a_0=p_1,\quad b_0=q_1,
\end{equation}
and again impose the equations of motion at $\alpha_ih$, and so on.
\end{enumerate}
How are the $\alpha_i$'s determined?  By requiring preservation of
the
canonical commutation relations at each lattice site,
\begin{equation}
[q_1,p_1]=[q_0,p_0]=i,
\label{unitarity}
\end{equation}
one finds
\begin{eqnarray}
r=1 \quad (\mbox{linear finite elements}) \qquad \alpha&=&{1\over2}\\
r=2 \quad (\mbox{quadratic finite elements}) \qquad
\alpha_\pm&=&{1\over2}\pm
{1\over2\sqrt{3}}\\
r=3 \quad (\mbox{cubic finite elements}) \qquad
\alpha_{1,3}&=&{1\over2}\mp{\sqrt{3}\over
2\sqrt{5}},\quad \alpha_2={1\over2}
\end{eqnarray}
These points are exactly the Gaussian knots, that is, the roots of
the
$r$th Legendre polynomial,
\begin{equation}
P_r(2\alpha-1)=0.
\end{equation}
Amazingly, these are precisely the points at which the numerical
error
is minimized.  It is known for classical equations that if one uses
$N$ $r$th degree finite elements the relative error goes like
$N^{-2r}$,
while imposing the equations at any other points would give errors
like $N^{-r}$.

Let us consider a simple example.  The quartic anharmonic oscillator
has continuum Hamiltonian
\begin{equation}
H={1\over2} p^2+{1\over4}\lambda q^4,
\end{equation}
for which the equations of motion are
\begin{equation}
\dot q=p\quad \dot p=-\lambda q^3.
\end{equation}
If we use the linear ($r=1$) finite-element prescription given above,
the corresponding discrete lattice equations are
\begin{equation}
{q_1-q_0\over h}={p_1+p_0\over2},\quad {p_1-p_0\over
h}=-{\lambda\over8}
(q_1+q_0)^3.
\label{leqm}
\end{equation}
(Notice the easily remembered mnemonic for linear finite elements:
Derivatives are replaced by forward differences, while
undifferentiated
operators are replaced by forward averages.)
By commuting the first of these equations with $p_1+p_0$ and the
second
with $q_1+q_0$ the unitarity condition (\ref{unitarity}) follows
immediately.  These equations are implicit, in the sense that we
must solve a nonlinear equation to find $q_1$ and $p_1$ in terms
of $q_0$ and $p_0$.  Although such a solution can be given, let
use make a simple approximation, by expanding the dynamical operators
at time 1 in powers of $h$, with operator coefficients at time 0.
Those coefficients are determined by (\ref{leqm}), and a very simple
calculation yields
\begin{eqnarray}
q_1&=&q_0+hp_0-{\lambda\over2}h^2 q_0^3+\dots,\nonumber\\
p_1&=&p_0-\lambda h q^3_0-{3\over2}\lambda
h^2q_0p_0q_0+\dots.\label{exp}
\end{eqnarray}
We can define Fock space creation and annihilation operators in
terms of the initial-time operators
\begin{equation}
q_0=\gamma{(a+a^\dagger)\over\sqrt{2}},\quad p_0={(a-a^\dagger)\over
i\sqrt{2}
\gamma},
\label{aadag}
\end{equation}
which satisfy
\begin{equation}
[a,a^\dagger]=1.
\end{equation}
Here we have introduced an arbitrary variational parameter $\gamma$.
The Fock-space states (harmonic oscillator states) are created and
destroyed
by these operators:
\begin{equation}
|n\rangle={(a^\dagger)^n\over\sqrt{n!}}|0\rangle,
\label{hostates}
\end{equation}
which states are not energy eigenstates of the anharmonic oscillator.
We can now take matrix elements in these states of the dynamical
operators
at lattice site 1, using (\ref{exp}):
\begin{eqnarray}
\langle 1|p_1|0\rangle&\approx&\langle 1|p_0|0\rangle
(1+i{3\over2} h\lambda
\gamma^4-{3\over4}h^2\lambda\gamma^2+\dots)\nonumber\\
&\approx&\langle 1|p_0|0\rangle(1+i\omega h-{1\over2}\omega^2
h^2+\dots),
\end{eqnarray}
and
\begin{eqnarray}
\langle 1|q_1|0\rangle&\approx&\langle 1|q_0|0\rangle
(1+i{h\over\gamma^2}  -{3\over4}h^2\lambda\gamma^2+\dots)\nonumber\\
&\approx&\langle 1|q_0|0\rangle(1+i\omega h-{1\over2}\omega^2
h^2+\dots),
\label{q1}
\end{eqnarray}
where we have assumed approximately exponential dependence
on the energy difference $\omega$.  Equating the coefficients
of the terms through order $h^2$ constitutes four
equations in two unknowns. These equations are consistent and yield
\begin{equation}
\omega={3\over2}\lambda\gamma^4={1\over\gamma^2},
\end{equation}
so the energy difference between the ground state and the first
excited
state is approximately
\begin{equation}
\omega=\left({3\over2}\lambda\right)^{1/3}\approx1.145\lambda^{1/3}
\end{equation}
which is only 5\% higher than the exact result
$E_{01}=1.08845\lambda^{1/3}$.
A similar calculation using quadratic finite elements ($r=2$) reduces
the
error to 0.5\%.

\section{The Time-Evolution Operator}
Because the canonical commutation relations are preserved at each
lattice
site, we know that there is a unitary time evolution operator that
carries dynamical variable forward in time:
\begin{equation}
q_{n+1}=Uq_nU^{\dagger},\quad p_{n+1}=Up_nU^\dagger,
\end{equation}
 For the system described by the continuum Hamiltonian (\ref{ham}) in
the
linear finite-element scheme, we have found \cite{dtqm}
the following formula for
$U$:
\begin{equation}
U=e^{ihp_n^2/4}e^{ihA(q_n)}e^{ihp_n^2/4},
\end{equation}
where
\begin{equation}
A(x)={2\over h^2}[x-g^{-1}(4x/h^2)]^2+V(g^{-1}(4x/h^2)),
\end{equation}
\begin{equation}g(x)={4\over h^2}x+V'(x).
\label{gee}
\end{equation}
The implicit nature of the finite-element prescription is evident
in the appearance of the inverse of the function $g$.

Given the time evolution operator, a lattice Hamiltonian may be
defined
by $U=\exp(ih{\cal H})$.  For linear finite elements ${\cal H}$
differs
from the continuum Hamiltonian by terms of order $h^2$.  For example,
\begin{eqnarray}
V={1\over2}m^2q^2: & & {\cal H}={2\over
mh}\tan^{-1}\left(mh\over2\right)
\left[{1\over2}p^2+{1\over2}m^2q^2\right],\label{hc}\\
V={\lambda\over3}q^3: & & {\cal H}={1\over2}p^2+{1\over3}\lambda q^3
+h^2\left[{\lambda\over12}pqp+p^3\right]+\dots,\\
V={\lambda\over4}q^4: & & {\cal H}={1\over2}p^2+{1\over4}\lambda q^4
+h^2\left[-{\lambda^2\over24}q^6-{\lambda\over8}qp^2q\right]+\dots.
\label{ahc}
\end{eqnarray}
If one uses quadratic finite elements ${\cal H}$ differs from the
continuum Hamiltonian by terms of order $h^4$, etc.

\section{Matrix Elements of Dynamical Variables}
\label{matq}
Remarkably, it is not necessary to solve the equations of motion
to compute matrix elements of the dynamical variable.  Introduce
creation and annihilation operators as in (\ref{aadag}).  Then, in
terms
of harmonic oscillator states (\ref{hostates}) the following formula
is
easily derived \cite{bss} for a general matrix element of $q_1$:
\begin{eqnarray}
\langle
m|q_1|n\rangle&=&-{\gamma\over\sqrt{2}}(\sqrt{m}\delta_{n,m-1}
+\sqrt{n}\delta_{m,n-1})\nonumber\\
+{e^{-i\theta(m-n)}\over R\sqrt{\pi
2^{n+m}n!m!}}&&\int_{-\infty}^\infty
dz\,ze^{-g^2(z)/4R^2}g'(z)H_n(g(z)/2R)H_m(g(z)/2R),
\end{eqnarray}
where $g$ is given by (\ref{gee}), $H_n(x)$ is the $n$th Hermite
polynomial,
 and we have introduced the abbreviations
\begin{equation}
R^2={4\gamma^2\over h^4}+{1\over h^2\gamma^2},\quad e^{-i\theta}
={2\gamma\over Rh^2}+{i\over R h\gamma}.
\end{equation}
For the example of the harmonic oscillator, this formula gives for
the
ground state--first excited state energy difference $\omega=(2/ h)
\tan^{-1}(h/2)$, consistent with (\ref{hc}), while for the anharmonic
oscillator if we expand in $h$ we obtain precisely the expansion
(\ref{q1}).

\section{Matrix Elements of the Time Evolution Operator}
\label{matu}
A similar formula can be derived for the harmonic oscillator matrix
elements of the time evolution operator.  (There is an error in the
formula printed in \cite{bss}.)
\begin{eqnarray}
\langle m|U|n\rangle&=&{1\over2R}{1\over\sqrt{\pi
2^{n+m}n!m!}}e^{-i(n+m+1)\theta}\nonumber\\
&\times&\int_{-\infty}^\infty
dz\,g'(z)H_n(g(z)/2R)H_m(g(z)/2R)e^{[ihV(z)+ih^3V'(z)^2/8
-h^2g(z)^2e^{-i\theta}/8\gamma R]},
\label{umat}
\end{eqnarray}
which again is expressed in terms of $g$ not $g^{-1}$.

For the harmonic oscillator, where $V=q^2/2$, (\ref{umat}) gives for
the
ground-state energy
\begin{equation}
\langle 0|U|0\rangle=e^{i\omega_0h},\quad \omega_0=
{1\over h}\tan^{-1}{h\over2},
\end{equation}
which follows from (\ref{hc}).  For the anharmonic oscillator,
$V=\lambda q^4/4$,
again, for a first look,
 we expand in powers of $h$, with the result, for the harmonic
oscillator ground state,
\begin{eqnarray}
\langle
0|U|0\rangle&=&1+h\left({i\over4\gamma^2}+{3i\over16}\lambda\gamma^4
\right)+h^2\left(-{3\over32\gamma^4}+{9\over64}\lambda\gamma^2
-{105\over512}\lambda^2\gamma^8\right)+\dots\nonumber\\
&\approx&1+i\omega_0h-{1\over2}\omega^2_0h^2+\dots,
\label{woo}
\end{eqnarray}
which is also derivable from (\ref{ahc}).
Equating powers of $h$ gives us two equations, which
 are to be solved first for the dimensionless number
$\lambda\gamma^6=\alpha$.
Once the number $\alpha$ is determined, the value of $\omega_0$ is
expressed as
\begin{equation}
\omega_0=\lambda^{1/3}f(\alpha),\quad f(\alpha)={1\over4\alpha^{1/3}}
\left(1+{3\over4}\alpha\right).
\end{equation}
%The function $f(\alpha)$ is shown in Fig.~\ref{fig:figa}:
%\begin{center}

%\begin{figure}[ht]
%   \centerline{\psfig{figure=figa.ps,height=6.5in,width=5in}}
 %  \caption{Ground-state energy for the anharmonic
%oscillator as a function of
%$\alpha=\lambda\gamma^6$, in the first approximation}
 %  \label{fig:figa} %Label to reference in text
   %(Figure~\ref{fig:filename})
%\end{figure}

%\end{center}
For a first estimate, we use the ``principle of minimum
sensitivity'', that
is, use the stationary value of $\alpha$,
\begin{equation}
f'(\alpha)=0\Rightarrow \alpha={2\over3}\Rightarrow f(\alpha)=0.4293,
\end{equation}
which is about 2\% higher than the exact value of 0.42081 \cite{bms}.
  In fact,
when we solve (\ref{woo}) for $\alpha$ we find a complex value
\begin{equation}
\alpha={1\over2}\pm{i\over2\sqrt{3}}\Rightarrow
f(\alpha)=0.4178\mp0.0077i.
\label{wo1}
\end{equation}
The imaginary part is small, and the real part is only 0.7\% low.
The failure of (\ref{wo1}) to be real does not indicate any breakdown
of unitarity, but only that the one state approximation is not exact.
We do much better by making a two-state approximation, where we must
diagonalize the $2\times2$ matrix
\begin{equation}
\left(\begin{array}{cc}
U_{00}&U_{02}\\
U_{20}&U_{22}\\
\end{array}\right).
\label{u}
\end{equation}
We then find the following relation between $\omega_{0,2}$ and
$\alpha=\lambda\gamma^6$:
\begin{equation}
\omega_{0,2}={\lambda^{1/3}\over16}\alpha^{-1/3}[12+21\alpha
\mp2\sqrt{3}(8+16\alpha+33\alpha^2)^{1/2}],
\label{omgood}
\end{equation}
which, for the $-$ sign, is plotted in Fig.~1:
 %and \ref{fig:figc}.
%\begin{center}

\begin{center}
  \centerline{\psfig{figure=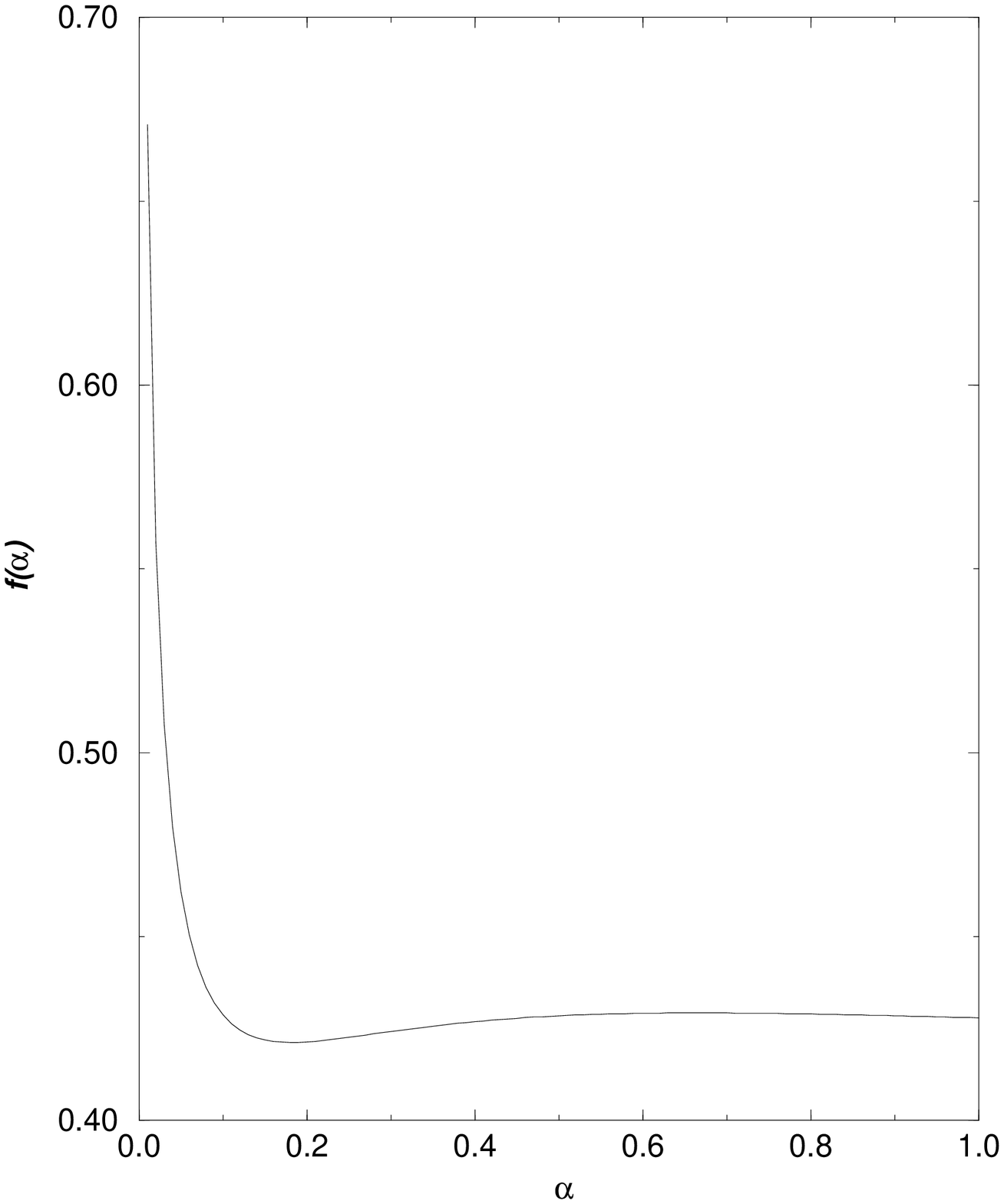,height=6.5in,width=5in}}
 %\label{fig:hofig.ps}
 \begin{quotation} FIG.~1. Ground-state energy for the anharmonic
oscillator as a function of
$\alpha=\lambda\gamma^6$, in the second approximation.
Here $\omega_0=\lambda^{1/3}f(\alpha)$, $f(\alpha)$ given by
(\ref{omgood}).
\end{quotation}
   %Label to reference in text
%  (Figure~\ref{fig:filename})
\end{center}
%\end{center}
%\begin{center}

%\begin{figure}[ht]
%   \centerline{\psfig{figure=figc.ps,height=6.5in,width=5in}}
%   \caption{Second excited state energy for the anharmonic
%oscillator as a function of
%$\alpha=\lambda\gamma^6$, in the second approximation}
 %  \label{fig:figc} %Label to reference in text
%(Figure~\ref{fig:filename})
%\end{figure}
%\end{center}

This  graph shows that the ground-state energy is very insensitive
to the value of $\alpha$.  The principle of minimum sensitivity
give spectacular agreement with the exact result,
\begin{equation}
\omega_0=0.421235\lambda^{1/3},
\end{equation}
being only 0.1\% high,
while it gives a good value for the third state,
$\omega_2=2.992\lambda^{1/3}.$
Solving for $\alpha$ from the eigenvalues of (\ref{u}) gives even
better results:
\begin{equation}
\omega_0=\lambda^{1/3}(0.42054+2\times 10^{-6}i), \quad
\omega_2=\lambda^{1/3}(2.94328-.022029i),
\end{equation}
where the ground state energy is now low by 0.06\%, the imaginary
part
being negligible.

\section{Conclusions}

The simple calculations given here for the quantum-mechanical
anharmonic oscillator are the beginning of a program to develop
use of lattice Hamiltonian techniques to explore gauge theories
in the finite-element context. The astute reader will note
that the numerical results presented in Sec.~\ref{matu}
also hold in the continuum, by virtue of (\ref{ahc}).  It is in
two or more space-time dimensions that the essential nature of
the lattice in such calculations comes into play
\cite{mmsb,mmsb2,bm}.
 The high accuracy contrasted with
the simplicity of the approach leads us to expect that we can
extract spectral information, anomalies, and symmetry breaking
from an examination of the time-evolution operator.

%\begin{quotation}
%FIG. 1.  Caption of Figure 1 (arabic num.).  The figure caption
%%should
%be placed below the figure and should be sequentially numbered.  The
%figure and its caption should be placed on the same page.
%\end{quotation}

\section*{Acknowledgments}
I thank the US Department of Energy for partial financial support of
this
research.  I am grateful to Carl Bender, Pankaj Jain, and Dean Miller
for
useful conversations.

\end{document}